# BEAM LOSSES DUE TO THE FOIL SCATTERING FOR CSNS/RCS*

M.Y. Huang#, N. Wang, S. Wang, S.Y. Xu
Institute of High Energy Physics, Beijing, China

*Abstract*
For the Rapid Cycling Synchrotron of China Spallation Neutron Source (CSNS/RCS), the stripping foil scattering generates the beam halo and gives rise to additional beam losses during the injection process. The interaction between the proton beam and the stripping foil was discussed and the foil scattering was studied. A simple model and the realistic situation of the foil scattering were considered. By using the codes ORBIT and FLUKA, the multi-turn phase space painting injection process with the stripping foil scattering for CSNS/RCS was simulated and the beam losses due to the foil scattering were obtained.

## INTRODUCTION

CSNS is a high power proton accelerator-based facility [1]. The accelerator consists of an 80MeV H⁻ linear accelerator which is upgradable to 250MeV and a 1.6GeV RCS which accumulates an 80MeV injection beam, accelerates the beam to the designed energy of 1.6GeV and extracts the high energy beam to the target. Its beam power is 100kW and capable of upgrading to 500kW. The design goal of CSNS is to obtain the high intensity, high energy proton beam with a repetition rate of 25Hz for various scientific fields [2].

For the high intensity proton accelerators, injection via H⁻ stripping is actually a practical method. In order to control the strong space charge effects which are the main causes of the beam losses in CSNS/RCS, the phase space painting method is used for injecting the beam of small emittance from the linear accelerator into the large ring acceptance [3].

Beam losses in an accelerator can be divided into two categories: controlled beam losses and uncontrolled beam losses [4]. In high power accelerator, one potential source of the uncontrolled beam losses is the stripping foil scattering of the H⁻ ions. When the H⁻ beam traverses the stripping foil, most of the particles H⁻ are converted to H⁺, and the others are converted to H⁰ or unchanged [5][6]. The interaction of the H⁻ beam with the stripping foil can induce the beam losses. By using the codes ORBIT [7] which is a particle tracking code for rings and FLUKA [8] which is a code for calculations of particle transport and interactions, the multi-turn phase space painting injection process with the stripping foil scattering for CSNS/RCS can be simulated and the beam losses due to the foil scattering can be calculated.

## FOIL SCATTERING

Change exchange phenomena gives rise to capture or loss of electrons by the fast moving ions traverse a material [6]. For CSNS, when the H⁻ beam traverses the carbon stripping foil, there are six charge exchange processes: three are electron loss reactions and three are electron pickup reactions [9]. For energies above 100keV, the cross sections for electron pickup are very small and can be neglected. Therefore, the remained particles after foil stripping are H⁻, H⁰ and H⁺, as shown in Fig. 1.

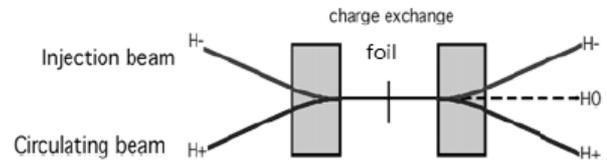

Figure 1: Stripping foil scattering.

During the injection process, the foil scattering can generate the beam halo and result in additional beam losses. The stripping foil scattering had been studied in detail for J-PARC and it can be found that the foil scattering was the main cause of beam losses in the injection region [10]. Therefore, the beam losses due to the foil scattering for CSNS also need to be studied in detail. Table 1 shows the beam parameters for 80MeV injection and 250MeV injection.

Table 1: Beam Parameters for 80 MeV Injection and 250 MeV Injection

| Injection | 80MeV | 250MeV |
|---|---|---|
| Injection beam power/kW | 5 | 80 |
| Average injection current/μA | 62.5 | 312.5 |
| Turn number of injection | 200 | 403 |
| Foil thickness/(μg/cm²) | 100 | 240 |

## A SIMPLE MODEL OF THE FOIL SCATTERING

In order to obtain a preliminary judgment of the beam losses due to the foil scattering, in this section, a very simple model of the stripping foil scattering is adopted, and then simulated by the code FLUKA. A preliminary result of the beam losses can be obtained. The simple model will be used for J-PARC to check its accuracy firstly, and then used for CSNS to calculate the beam losses in the injection region.

Suppose the particle beam is uniform distribution and hitting on the stripping foil vertically. The detector is located near the stripping foil. The turn number that the particle beam traverses the stripping foil is considered simple as the average traversal number. Therefore, the total beam losses are the result of the beam losses in single turn multiplying by the average traversal number. By using the code FLUKA, the foil scattering can be simulated and the total beam losses can be obtained.

___________________________________________
*Work supported by National Natural Science Foundation of China (Project Nos. 11175020 and 11175193)
#huangmy@ihep.ac.cn

For J-PARC, by using the above simple model, the stripping foil scattering process is simulated. It can be found that the beam losses are about 1.6W for 181MeV injection and 9.3W for 400MeV injection. In order of magnitude, this simulation result is consistent with that given by J-PARC [10]. Therefore, the simple model is correct in order of magnitude and can be used for CSNS to give a preliminary judgment of the beam losses.

For CSNS, by using the above simple model, the stripping foil scattering process is also simulated and the results are given in Table 2. It can be found that the beam losses are about 0.1W for 80MeV injection and 2.2W for 400MeV injection. Therefore, a preliminary result of the beam losses is obtained.

Table 2: Beam Losses due to the Simple Model of the Foil Scattering

| Injection | 80MeV | 250MeV |
|---|---|---|
| Average traversal number | 5 | 10 |
| Particle loss ratio in one turn | 0.00025% | 0.00028% |
| Total beam losses/W | 0.1 | 2.2 |

## REALISTIC FOIL SCATTERING

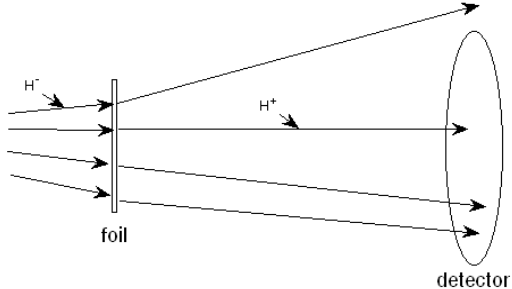

Figure 2: Detection of the realistic foil stripping.

In the above section, a simple model of the foil scattering was adopted and a preliminary result of the beam losses was obtained. In this part, the realistic foil scattering is simulated by the codes ORBIT and FLUKA, and the accurate result of the beam losses will be obtained.

Considering the particle beam distribution which given by the simulation of the multi-turn phase space painting injection process. The detector is located at the nearest quadrupole and its size is the aperture of the quadrupole. The turn number that the particle beam traverses the stripping foil is considered simple as the average traversal number. Figure 2 shows the detection of the realistic foil scattering.

For CSNS, by using the code ORBIT, the multi-turn phase space painting injection process is simulated, and the average traversal number and the beam distribution after injection are obtained. Calculating those particles of the beam distribution which are in the range of the stripping foil, the twiss parameters and 99% emittance for those particles are obtained, as shown in Table 3. With these beam parameters, the beam distribution that hitting on the stripping foil can be fitting. Then, the foil scattering process can be simulated by the code FLUKA and the beam losses due to the foil scattering in single turn are obtained. By using the average traversal number, the foil scattering induced beam losses during the multi-turn injection process can be calculated. Table 4 shows a summary of the beam losses due to the foil scattering. It can be found that the beam losses due to the realistic foil scattering are about 0.3W for 80MeV injection and 4.6W for 250MeV injection.

Table 3: Beam Parameters of the Proton Distribution that Hitting on the Stripping Foil

| Injection | 80MeV | 250MeV |
|---|---|---|
| $(\varepsilon_{x,99\%}, \varepsilon_{y,99\%})$/(mm · mrad) | (92, 247) | (90, 282) |
| $(\alpha_x, \alpha_y)$ | (0.003, 0.044) | (0.001, 0.016) |
| $(\beta_x, \beta_y)$/m | (1.833, 4.458) | (1.877, 5.222) |
| $(\gamma_x, \gamma_y)$/m$^{-1}$ | (0.546, 0.225) | (0.533, 0.192) |

Table 4: Beam Losses due to the Realistic Foil Scattering

| Injection | 80MeV | 250MeV |
|---|---|---|
| Average traversal number | 5 | 10 |
| Particle loss ratio in one turn | 0.0012% | 0.00058% |
| Total beam losses/W | 0.3 | 4.6 |

## CONCLUSIONS

For CSNS, during the injection process, the stripping foil scattering generates the uncontrolled beam losses. The interaction between the H$^-$ beam and the stripping foil was discussed and the foil scattering was studied. Firstly, a simple model of the foil scattering was adopted and a preliminary result of the beam losses was obtained. Then, the realistic foil scattering was simulated and the accurate result of the beam losses was obtained. It can be found that the beam losses due to the realistic foil scattering are about 0.3W for 80MeV injection and 4.6W for 250MeV injection.

## ACKNOWLENDGMENTS

The authors want to thank CSNS colleagues for the discussion and consultations.

# REFERENCES

[1] S. Wang et al., Chin Phys C, 33 (2009) 1-3.
[2] J. Wei et al., Chin Phys C, 33 (2009) 1033-1042.
[3] J.Y. Tang et al., Chin Phys C, 30 (2006) 1184-1189.
[4] R.C. Webber and C. Hojvat, IEEE Trans. Nucl. Sci. NS-26 (1979) 4012-4014.
[5] A.H. Mohaghegi et al., Phys. Rev. A, 43 (1991) 1345-1365.
[6] W. Chou, M. Kostin, and Z. Tang, Nucl. Instrum. Methods. Phys. Res. A, 590 (2008) 1-12.
[7] J. Gabambos et al., ORBIT User's Manual, SNS/ORNL/AP Technical Note 011. 1999.
[8] A. Ferrari et al., FLUKA: multi-particle transport code, CERN-2005-010, 2008.
[9] M.S. Gulley et al., Phys. Rev. A 53 (1996) 3201-3210.
[10] N. Akasaka et al., Operation software for commissioning of KEKB linac programmed with SAD, Proceedings of APAC, 1998. 495-497